\documentclass[aps,pra,twocolumn,showpacs]{revtex4}

\usepackage{amsmath}    
\usepackage{hyperref}   
\usepackage{amssymb}    
\usepackage{graphicx}   
\usepackage{epsfig}
\usepackage{verbatim}   
\usepackage{color}      

\newcommand{\bk}{\mathbf{k}}
\newcommand{\br}{\mathbf{r}}

\begin{document}
\title{Mean-field dynamics to negative absolute temperatures in the Bose-Hubbard model}
\author{\'Akos Rapp}
\affiliation{ Institut f\"ur Theoretische Physik, Leibniz Universit\"at, 30167 Hannover, Germany}

\date{\today}

\begin{abstract}
We apply time-dependent Gutzwiller mean-field theory to provide a qualitative understanding for bosons in optical lattices that approach states corresponding to negative absolute temperatures. We perform the dynamical simulations to relate to the recent experiments by Braun \textit{et al.} [[ S. Braun, J. P. Ronzheimer, M. Schreiber, S. S. Hodgman, T. Rom, I. Bloch and U. Schneider Science \textbf{339} 52 (2013)]]. Time-of-flight images calculated from the two-dimensional numerical simulations reproduce characteristics of the experimental observations, in particular, the emergence of the four peaks at the corners of the Brillouin zone.
\end{abstract}


\maketitle

\section{Introduction}

Physical systems with a perfect isolation and an upper bound on the energy are allowed to be in equilibrium states that are described by a negative absolute temperature, $T<0$.
These conditions are not too severe, yet such systems are only rarely observed. The main reason is that most systems are described by Hamiltonians that have no upper bounds: free particles with mass $M>0$ have kinetic energies $\mathbf{p}^2/2M$, or, using a more specific example, ultracold atoms are trapped by potentials $V(\mathbf{r}) \approx V_0 \, \mathbf{r}^2$ with $V_0 > 0$. 
Negative $T$ was first observed in nuclear spin systems ~\cite{nuclspins1,nuclspins2}, which have very weak coupling to the environment and naturally possess both lower and upper bounds on the energy due to the finite-dimensional Hilbert space of the spins.
It is much harder to achieve negative absolute temperatures for systems with motional degrees of freedom.
Nevertheless, ultracold atoms in optical lattices hold the possibility to observe this novel state of matter~\cite{Mosk,negT}, since in these atomic clouds the system parameters can be controlled experimentally with great flexibility and accuracy~\cite{coldatom-revmod}.

In a recent experiment, Braun \textit{et al.}~\cite{exp-negT} indeed showed that attractively interacting bosonic atoms in an antitrapping potential with $V_0 < 0$ can relax to a thermodynamically stable state with a macroscopic occupation [Bose-Einstein condensate (BEC)] of the highest-energy single-particle states. This is also a demonstration of the inverse population of energy levels, a striking property of systems with $T<0$, which can lead to counterintuitive phenomena.
Nevertheless, ultracold atomic experiments always need to start at $T>0$ and $V_0 > 0$, and a certain protocol (see, e.g., Ref.~\cite{negT}) is required to reach the $V_0 < 0$ side. Although numerical simulations for two-component fermionic clouds~\cite{negT} were performed to have an understanding about the dynamics as the $T<0$ state is approached, bosons were not yet studied in a time-dependent simulation. In this work we address this question to complement the results for fermions and, more importantly, for comparison with the current experiments.

In Sec. II, we discuss the experimental setup and the corresponding microscopic model. Section III is dedicated to the introduction and description of the numerical procedure. In Sec. IV we overview the numerical results and compare them to results obtained in the experiments. In Sec. V we discuss key aspects of the dynamics of the cloud at different stages of the simulations, including dephasing in the initial deep lattice, the melting rate of the Mott region during the transient, and the spatial distribution of the condensate.

\section{Experimental setup and model}

We are modeling the experiments discussed in Ref.~\cite{exp-negT}, where $^{39}$K atoms are used to reach $T<0$, based on the protocol outlined below. In 25 ms, a three-dimensional optical lattice at laser wavelength $\lambda_L = 736.65$ nm is ramped up to a lattice depth $s =22 E_R$, where $E_R = \hbar^2 k_L^2/(2M) $ is the recoil energy and $k_L=2\pi/\lambda_L$. Using this deep lattice, in combination with a large positive scattering length $a_s=309 a_{\rm Bohr}$ and a tight harmonic confinement, most of the cloud is in a Mott phase with one atom per lattice site. In the next step, in 2 ms, the magnetic field is ramped over a Feshbach resonance to reach a negative scattering length, $a_s<0$, and simultaneously  the red-detuned dipole traps in the horizontal directions are also ramped down. Thus the blue-detuned optical lattice beams provide an effective antitrapping potential, $V_0 < 0$. After a waiting period of 1 ms, the horizontal optical lattice beams are ramped down to the final lattice depth, $s_{{\rm hor},f} =6 E_R$, in 2.5 ms.
After this point ($t=30.5$ ms), the cloud is kept for various times with the final experimental parameter values and the coherence of the condensate is analyzed based on time-of-flight images.
Due to technical difficulties regarding the compensation of gravity and to reach strong antitrapping potentials, the vertical optical lattice is kept at a depth of $s_{\rm ver} =22 E_R$, and the tight vertical confinement is never reversed. We use this detail to our advantage to greatly simplify the numerical simulations.

With the assumption that the atoms are confined to the lowest Bloch band of the lattice~\cite{coldatom-revmod}, they can be described by the Bose-Hubbard model~\cite{BHM-1,BHM-2}, defined by
\begin{eqnarray}
 H &=& -J\!\sum_{\langle ij\rangle}\!b_i^\dagger b_j^{\phantom{\dagger}}
    +\sum_j\!\left[ U \frac{\hat{n}_j-1}{2} +  V_0 \, \mathbf{r}_j^2 - \mu_0 \right] \hat{n}_j , \label{eq:BHM}
\end{eqnarray}
where $\langle ij\rangle$ denotes nearest neighbors, $J$ is the nearest-neighbor hopping rate, $V_0$ characterizes the strength of the harmonic potential, $\mu_0$ is a chemical potential to fix the number of atoms, and $U$ is the on-site interaction strength. Bosons are created and destroyed at a site $j$ by $b_j^{\dagger}$ and $b_j^{\phantom{\dagger}}$, respectively, and $\hat{n}_j = b_j^{\dagger}b_j^{\phantom{\dagger}} $.

Approximately $1\times10^5$ atoms were used in the three-dimensional setup of the experiment. We simplify our numerical calculations and consider only a two-dimensional horizontal layer of the atomic cloud. The simulations start at $t=20$ ms of the experiment. At this time, the lattice depth is already quite deep, $s\approx17.6 E_R$, and we neglect the vertical hopping between the layers.

For simplicity, we calculate the hopping $J$ and interaction strength $U$ in harmonic approximation, where they depend on the lattice depths and the magnetic field $B$ as~\cite{coldatom-revmod}
\begin{equation}
 J(s_{\rm hor})/E_R = \frac{4}{\sqrt{\pi}} s_{\rm hor}^{3/4} e^{-2 \sqrt{s_{\rm hor}}}, \label{eq:J}
\end{equation}
and 
\begin{equation}
 U(s_{\rm hor},s_{\rm ver}, B)/E_R = \sqrt{\frac{8}{\pi}} \, k_L a_s(B) \, s_{\rm hor}^{2/4} s_{\rm ver}^{1/4}. \label{eq:U}
\end{equation}
In the case of the interaction, the anisotropy of the Wannier functions in the horizontal and vertical directions has been taken into account.
For $^{39}$K atoms the scattering length as a function of the magnetic field $B$ near the Feshbach resonance used in the experiment is approximately~\cite{K39-Feshbach} 
\begin{equation}
 a_s(B) = a_{\rm bg} \left[ 1 - \Delta/(B - B_{\rm res}) \right], \label{eq:as}
\end{equation}
where $a_{\rm bg} = -33  a_{\rm Bohr}$, $\Delta = -52$ G, and  $B_{\rm res} = 402.4$ G.

\section{Method}

To simulate the real-time dynamics of bosons, we apply time-dependent Gutzwiller mean-field theory. 
The bosonic Gutzwiller ansatz (GA) was first used to approximate the ground state of the homogeneous version of Eq.~(\ref{eq:BHM})~\cite{Gutzwiller-Kotliar,Gutzwiller-Krauth}. 
Later the approach was generalized to spatially inhomogeneous and also time-dependent situations and this time-dependent GA was
applied for various nonequilibrium scenarios recently~\cite{TDG-1,TDG-2,TDG-3,TDG-4,TDG-5,TDG-6}. 
Similar to the original GA, the time-dependent version also neglects terms beyond the leading order
in expansions in the inverse of the lattice coordination number~\cite{TDG-zinv}.
A rigorous connection to the mean-field limit was established for the time-dependent GA in Ref.~\cite{TDG-3}. 
In the time-dependent GA, the time evolution of the $m$-boson occupation amplitude at site $j$ of the lattice and time $t$, denoted by $f_m(j,t)$, is given by the following coupled differential equations:
\begin{eqnarray}
 i \partial_t f_m(j,t) &=& \left[U(t) \frac{(m-1)}{2} + V_0 (t) \, \mathbf{r}_j^2-\mu_0 \right]m \, f_m(j,t) \nonumber \\
 && - J(t) \; \phi^*(j,t) \; \sqrt{m+1} \; f_{m+1}(j,t) \nonumber \\
 && - J(t) \; \phi^{\phantom{*}}(j,t) \; \sqrt{m} \; f_{m-1}(j,t) .  
 \label{eq:tdG}
\end{eqnarray}
The mean field enters in the hopping term of Eq.~(\ref{eq:BHM}), as 
$\phi(j,t) = \sum_{\delta,m}\!\!\left[ \sqrt{m+1} \; f_m^*(j+\delta,t) f_{m+1}(j+\delta,t)  \right]$, where $\delta$ runs over the nearest neighbors of $j$, thus coupling the occupation amplitudes
at different sites. The amplitudes are normalized, $1 = \sum_m  |f_m(j ,t)|^2$, $\forall t,j$. 

The main focus of this work is the evolution of the dynamical system defined by Eq.~(\ref{eq:tdG}), where the parameters are changed in time according to various protocols. We also investigate the extent this relatively simple model captures the complex out-of-equilibrium dynamics of the quantum many-body system in the experiments.

The explicit time dependence of the microscopic parameters $U(t)$ and $J(t)$ is computed using Eqs. (\ref{eq:J}), (\ref{eq:U}) and (\ref{eq:as}) from the time-dependent optical lattice depths $s_{\rm hor}(t)$ and $s_{\rm vert}(t)$ and the magnetic field $B(t)$, while $V_0(t)$ is determined from the horizontal frequency using $ V_0(t) \sim \omega_{\rm hor}^2(t)$. More details can be found in Ref.~\cite{exp-negT}.

At $t_0=20$ ms, the initial condition $f_m(j,t_0)$ is determined in two steps. At each site the amplitudes are first calculated in GA in the local density approximation, followed by a few self-consistency sweeps  until the stationary solution of Eq.~(\ref{eq:tdG}) is found.
This corresponds to a tightly compressed Mott insulator with one atom per lattice site, surrounded by a tiny superfluid shell, the microscopic parameters being $J/U \approx 0.0023$ and $\mu_0/U \approx 0.15$. 
We used square lattices between $80\times 80$ and $160 \times 160$ lattice sites and an initial atom number of $N_{\rm tot}(t_0) \approx 1920$. The maximal allowed occupation number was usually $m_c=6$.

To numerically integrate Eq.~(\ref{eq:tdG}), we use a split-step method. At each time step $t \to t+\delta t$, we first perform unitary rotations corresponding to the diagonal terms in Eq.~(\ref{eq:tdG}). Second, we update the mean field $\phi$. Third, we perform an Euler step using the remaining terms. As the last step, we normalize the amplitudes.

Next, we analyze the results of the simulations and compare them to the experiments. The naming conventions of the various protocols and the corresponding final experimental parameters are summarized in Table~\ref{tbl:protocols}.

\begin{table}[ht]
 \caption{\label{tbl:protocols} Naming convention for the different protocols. The first parts of the protocols ($t < 25$ ms) are identical; the difference is in the final values of the interaction strengths and the external harmonic potentials. See also Figs.~\ref{fig:a0} and \ref{fig:cd}. }
\begin{tabular*}{0.48\textwidth}{lcc}
  \hline\hline
  Protocol name & $U_f \sim a_{s,f}$ & $V_{0,f} \sim (\omega_{{\rm hor},f}/2\pi)^2$ \\
  \hline
  a) & $-37 a_{\rm Bohr}$ & $- 43^2$ s$^{-2}$ \\
  
  c) & $-37 a_{\rm Bohr}$ & $+ 35^2 $ s$^{-2}$ \\
  
  d) & $-37 a_{\rm Bohr}$ & $+ 42^2 $ s$^{-2}$ \\
  
  e) & $-37 a_{\rm Bohr}$ & $+ 60^2 $ s$^{-2}$ \\
  
  z) & $+33 a_{\rm Bohr}$ & $+ 44^2$ s$^{-2}$\\
  \hline\hline
 \end{tabular*}
\end{table}

\section{Results}

\begin{figure}
 \includegraphics[width=0.45\textwidth,clip=true]{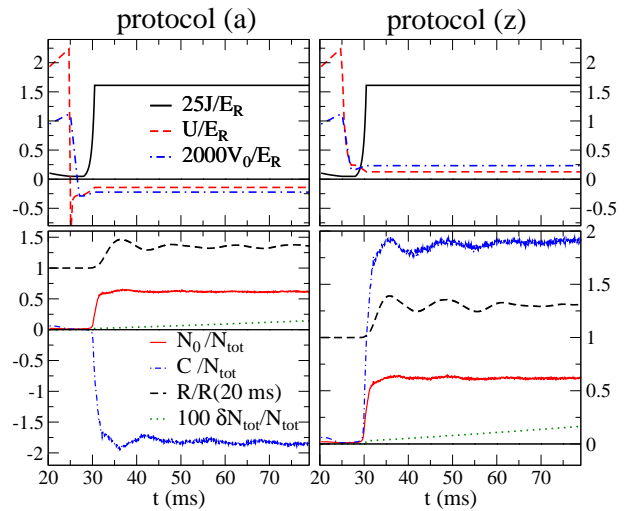}
 \caption{(color online) \label{fig:a0} Microscopic parameters (upper panels) and the corresponding macroscopic quantities (lower panels) from time-dependent Gutzwiller simulations for the protocols a) (left panels) and z) (right panels). The system approaches a macroscopically stationary state even for attractive final interactions $U_f<0$ and antitrapping potentials $V_{0,f}<0$.  
 We define $\delta N_{\rm tot}(t) = N_{\rm tot}(t)-N_{\rm tot}(20 {\rm ms})$. The noise [visible only for $N_0(t)$ and $C(t)$] is related to finite-size fluctuations.  
 }
\end{figure}

\begin{figure}
\centering
 \includegraphics[width=0.48\textwidth,clip=true]{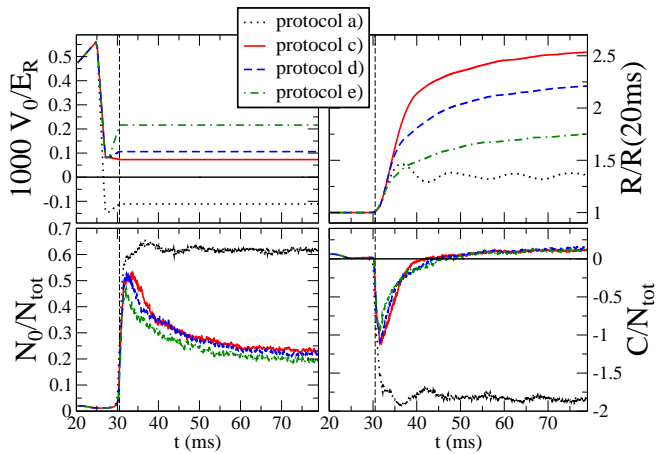}
 \caption{(color online) \label{fig:cd} Macroscopic quantities from time-dependent Gutzwiller simulations for the protocols c) (red solid line), d) (blue dashed line), and e) (green dash-dotted line). The system expands and does not relax to a stationary state. The SF coherence decays on short time scales. The vertical dashed lines correspond to $t=30.5$ ms. Data for protocol a) are shown for comparison (black dotted line). }
\end{figure}

We calculate various macroscopic quantities using the simulation, the total atom number
\begin{eqnarray}
 N_{\rm tot}(t) &=& \sum_{j} n_j(t),
\end{eqnarray} 
with $n_j(t) = \sum_m m |f_m(j,t)|^2$, the cloud radius
\begin{eqnarray}
 R(t) &=& \sqrt{\sum_{j} \mathbf{r}_j^2 \, n_j(t) / N_{\rm tot}(t)  } ,
\end{eqnarray}
the condensate fraction
\begin{eqnarray}
 N_{0}(t) &=& \sum_{j} |\langle b_j(t) \rangle|^2,
 \end{eqnarray}
with $\langle b_j(t) \rangle = \sum_m \sqrt{m+1} f_m^*(j,t) f_{m+1}(j,t)$, and 
(by adapting the terminology of Ref.~\cite{TDG-4}) the cloud average of 
the nearest-neighbor coherences
\begin{eqnarray}
 C(t) &=& \sum_{<ij>} \langle b_i^\dagger(t) b_j(t) \rangle \overset{\rm GA}{=} \sum_{<ij>} \langle b_i^\dagger(t) \rangle \langle  b_j(t) \rangle ,
\end{eqnarray}
where the last equation follows from the GA.
Note that $C$ is real and related to the kinetic energy, given by
\begin{equation}
 K(t) = - J(t) \, C(t) \;. \label{eq:KJC}
\end{equation}
We plot these quantities along with microscopic parameters for different protocols in Figs.~\ref{fig:a0} and \ref{fig:cd}. Although $N_{\rm tot}$ is conserved by Eq.~(\ref{eq:tdG}) ~\cite{TDG-1,TDG-2}, there is a weak time dependence due to errors of the numerical integration.
For $t>30.5$ ms, the time step was $\delta t = 0.1$ ns and the relative shift in the total atom number was less than 0.2\%, even at $t=80$ ms. Increasing the time step does not lead to a noticeable effect in the other macroscopic quantities.
We only note here that the noise in $N_0(t)$ and $C(t)$ is related to finite size fluctuations; we return to them later.

We see that when the signs of the final values of the interaction, $U_f$, and the harmonic potential, $V_{0,f}$, are the same, the system approaches a macroscopically stationary state. This is consistent both with the theoretical expectations that the cloud can equilibrate when the energy spectrum is bounded from at least one side, and with the experimental demonstration of the thermodynamic stability in these cases~\cite{exp-negT}. There is, however, a striking difference between the repulsively and the attractively interacting cases. In the more usual, repulsive case, with $U_f>0$ and $V_{0,f}>0$, the average coherence on nearest-neighbor sites is positive, $C>0$, and the system has a negative total kinetic energy. In contrast, in the attractive case, with $U_f < 0$ and $V_{0,f} < 0$ , the superfluid order parameter on nearest neighbors shows, on average, 
\emph{anticoherence}, $C < 0$, which directly translates to a \emph{positive} total kinetic energy [cf. Eq.~(\ref{eq:KJC})].
With the symmetric single-particle band and weak interactions, $K >0$ can only occur with an \emph{inverse population} of the single-particle energy levels. From a different perspective, the superfluid (SF) order parameter is alternating on the two sublattices, $\langle\hat b_j\rangle \approx (-1)^j \langle\hat b_0\rangle$ in the case a). Thus, the spatial Fourier transform (related also to time-of-flight images; see below) is enhanced around momenta $\mathbf{Q} =(\pm k_L,\pm k_L)$.
We note that the ground state of Eq.~(\ref{eq:BHM}) with parameters ${\rm sgn}(U_f) \, J, |U_f|$ and $|V_{0,f}|$ and the same $N_{\rm tot}$ as in the protocols a) or z) have superfluid fractions $N_{0}^{(0)}/N_{\rm tot} \approx 0.97 $ and cloud radii $R^{(0)}/R(20{\rm ms}) \approx 1.1 $ in GA. These values suggest that the stationary states are ``heated'' in comparison to the corresponding ground states.

Quite remarkably, in addition to the protocols which lead to thermodynamically stable clouds in experiments, the Gutzwiller approach also seems to capture qualitatively correctly the cases with attractive final interactions, $U_f <0$, and trapping potentials, $V_{0,f} > 0$. In these cases, the Hamiltonian is not bounded, and equilibrium is not possible at any finite $1/T \neq 0$. The dynamics is governed by two microscopic processes: the cloud is allowed to expand (see Fig.~\ref{fig:cd}), provided that the increase in potential energy is covered by a decrease in interaction energy (not shown); i.e., the atoms try to cluster. However, the expansion also leads to the dilution of the cloud, and, thus, the probability of clustering decreases. Therefore, the expansion gets slower and slower. 
This expansion is compatible with the experimental findings~\cite{exp-negT}, where atoms leaving the trap and distortions from residual nonharmonic terms in the potential were observed.
Unfortunately, without a precise knowledge of the continuity equations for the atom and energy densities, it is not possible to extract any long-time expansion laws analytically.

\begin{figure*}
  \centering
  \includegraphics[width=0.95\textwidth,clip=true,trim=0 24cm 0 0]{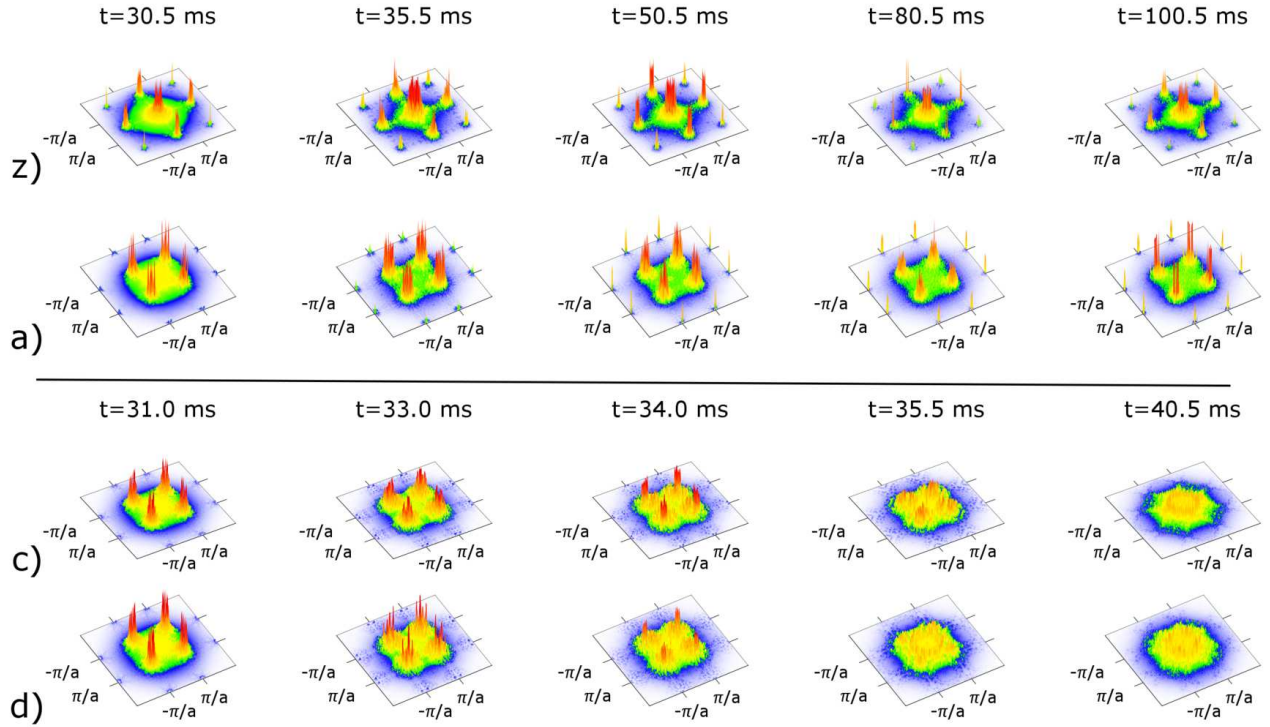}
  \caption{(color online) \label{fig:tof} Two-dimensional time-of-flight images from the Gutzwiller calculation for the different protocols at different times. The intensities $I_{\rm TOF}(\bk)$ have been normalized by the total particle number.  }
\end{figure*}

Finally, we focus on time-of-flight (TOF) images. Following Ref.~\cite{coldatom-revmod}, we define the (two-dimensional) TOF intensity as
\begin{equation}
 I_{\rm TOF}(\bk) \sim |w(\bk)|^2 {\cal G}(\bk)
\end{equation}
with the single-particle correlation function
\begin{equation}
 {\cal G}(\bk) \equiv \sum_{\br_i,\br_j} e^{i\bk(\br_i-\br_j)} \langle b^\dagger_i b^{\phantom{+}}_j \rangle \overset{\rm GA}{=} N_{\rm tot} - N_0  + |b(\bk)|^2 ,  \label{eq:Gk}
\end{equation}
where $ b(\bk) = \sum_{\br_j} e^{ - i\bk \br_j}  \langle  b_j \rangle $.
The Fourier transform of the Wannier function in harmonic approximation gives
$ |w(\bk)|^2 =   \frac{4 \pi}{\sqrt{s} k_L^2} \; e^{-\bk^2/(\sqrt{s} k_L^2) }. $
We plot TOF images in Gutzwiller approximation for the different protocols and different times in Fig.~\ref{fig:tof}, to be compared with the similar images in Ref.~\cite{exp-negT}. 

\begin{figure}
 \centering
 \includegraphics[width=0.45\textwidth,clip=true]{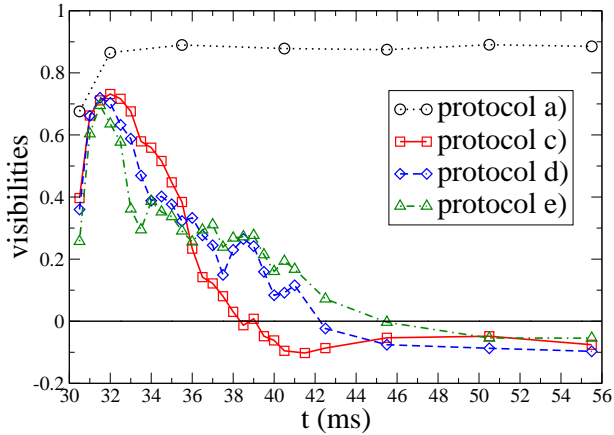}
 \caption{(color online) \label{fig:vis} Visibilities (defined similarly as in Ref.~\cite{exp-negT}) as a function of time $t$. For protocol a), we see no decay. For the protocols c), d), and e), the visibilities decay rapidly to negative final values, similar to the experiments.  }
\end{figure}

Based on the TOF images, we also define visibilities ${\cal V} = (I_A-I_B)/(I_A + I_B) $, with $I_{A,B}$ the intensities integrated in a small area around vectors $\mathbf{Q} = (k_L,k_L) $ and $\mathbf{\tilde Q} = (\sqrt{2} k_L,0)$, respectively. These are shown in Fig.~\ref{fig:vis}, and we see a fast decay to negative visibilities, similar to the experiments. However, we were not able to extract lifetimes, mainly due to the fluctuations in the data which can come from finite-size effects. 

\section{Discussion} 

After an overview of the numerical data and comparisons to the experiments, we discuss in more detail the dynamics of the atoms during the different stages of the protocols.

\begin{figure}
 \centering
 \includegraphics[width=0.48\textwidth,clip=true]{dephasing.eps}
 
 \includegraphics[width=0.48\textwidth,clip=true,trim=0 3cm 0 1cm]{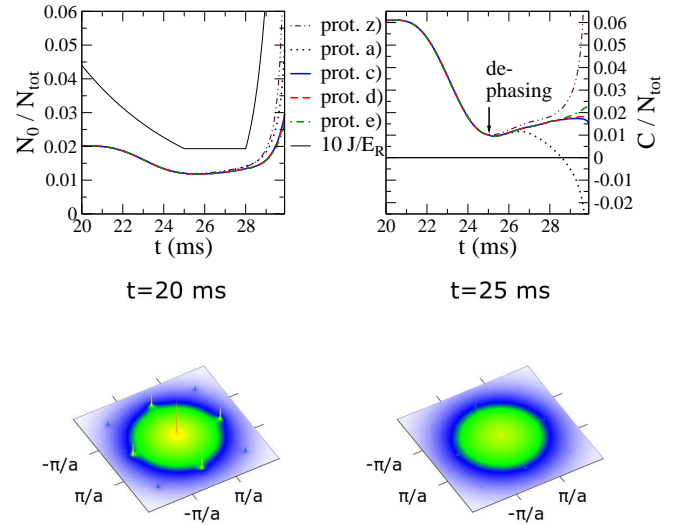}
 \caption{(color online) \label{fig:dephasing} Top: Initial evolution of the condensate fraction $N_0$ and nearest-neighbor coherences $C$. Dephasing of the initial weak condensate manifests in the substantially larger decrease in $C$ than in the $N_0$ between $t=20$ and $25$ ms. Lower panel: TOF images at $t=20$ and $t=25$ ms. The initial weak coherence peaks disappear completely by $t=25$ ms, also showing the dephasing effect. }
\end{figure}

For the initial part, $t<28$ ms, the cloud is in a tightly compressed Mott-insulating state. In the very deep optical lattice, the density distribution of the atoms is essentially frozen. Nevertheless, a tiny noninsulating (compressible) shell always exists around the Mott-insulating core. This region is initially a superfluid in the GA; however, it is subject to dephasing. Dephasing should help to optimize the final condensate fraction at $T<0$, as proposed in Ref.~\cite{negT}. To visualize this dephasing effect, we show the initial evolution of the condensate fraction $N_0$ and the nearest-neighbor coherences $C$ for $t< 30$ ms in Fig.~\ref{fig:dephasing}. In accordance with the still-increasing optical lattice depth, $N_0$ reaches a minimum around $t \approx 25$ ms, where its value is $\approx 60$\% of the initial value. In contrast, the $C$ decreases to lower than $20$\% of the initial value. 
While it is not possible to accurately quantify the dephasing, it is only partly responsible for the decrease. The effect can be understood relatively simply on the mean-field level: in the non
insulating region there is initially maximal coherence of $\langle b_j \rangle$  between different sites. 
However, phase differences accumulate due to the potential \emph{differences} between different sites: Eq.~(\ref{eq:tdG}) implies
\begin{eqnarray}
 -i \partial_t \langle b_j \rangle &=& (V_0 \mathbf{r}_j^2 - \mu_0) \langle  b_j \rangle  \\
	&+& U \sum_m \sqrt{m+1} \, m \, f_{m+1}^*(j) f_m(j)  + { O}(J).\nonumber
\end{eqnarray}
Here the second term is expected to be small as $n_j < 1$ and $|f_{m>1}| \approx 0$. 
As a consequence of the dephasing, the initial weak coherence peaks in the TOF images disappear completely by $t=25$ ms; see Fig.~\ref{fig:dephasing}.

\begin{figure}
 \centering
 \includegraphics[width=0.48\textwidth,clip=true]{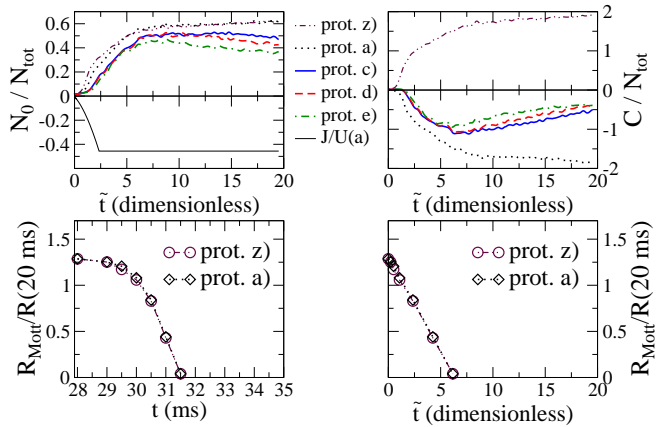}
 \caption{(color online) \label{fig:transient} Top: The transient in the condensate fraction $N_0$ (left) and nearest-neighbor coherences $C$ (right) in the natural time unit $\tilde t$, defined by Eq.~(\ref{eq:def:ttilde}). Bottom: Mott core radius $R_{\rm Mott}$ as a function of the times $t$ (left) and  $\tilde t$ (right). }
\end{figure}

Next, we concentrate on the transient period, $28 \leq t \leq 35$ ms. Since the transition from $N_0/N_{\rm tot} \approx 0.01$ to $N_0/N_{\rm tot} \approx 0.6$ happens quite rapidly, it is necessary to rescale the data as a function of the natural microscopic time scale instead of the laboratory time. Therefore, we show quantities in Fig.~\ref{fig:transient} as a function of the dimensionless time, defined by 
\begin{equation}
 \tilde t(t) := \int\limits_{28\;{\rm ms}}^t\!\!\!\!\frac{dt'}{\hbar} J(t') \;. \label{eq:def:ttilde}
\end{equation}
The sharp features of Figs.~\ref{fig:a0} and \ref{fig:cd} are indeed smoother in these units. One is tempted to compare features with quantum quenches from a Mott insulator to the superfluid with
 $U/J(28\; {\rm ms}) \approx -138.$,
 $U/J(30.5\; {\rm ms}) \approx -2.19$,
 and $\tilde t(30.5\; {\rm ms}) -\tilde t(28\; {\rm ms}) \approx 2.35$ .
However, we have to keep in mind that the state at $28$ ms is not the actual ground state due to the dephasing in the noninsulating region, as we discussed above.

It is interesting to measure how the Mott insulator melts during the transient in the GA. To address this question, we define the radius of the Mott insulator based on the region where the condensate fraction is negligible in the core of the cloud:
\begin{equation}
 R_{\rm Mott}(t) := \max\limits_{r} \; \lbrace r \; | \; 0.001 > \sum_{j:|\br_j|<r} |\langle  b_j(t)\rangle|^2  \rbrace \;. \label{eq:def:rMott}
\end{equation}
This radius is also shown in Fig.~\ref{fig:transient}, as a function of both real time $t$ and the natural time $\tilde t$. Interestingly, in the latter units, the radius of the Mott insulator decreases \emph{linearly}, i.e., the number of atoms in the Mott core decreases quadratically. It is unclear whether the constant speed $\partial R_{\rm Mott}/\partial \tilde t$ is related to the constant spreading velocity of off-diagonal long-range order found for quenches from the Mott insulator to the superfluid in the homogeneous case in Ref.~\cite{TDG-zinv}. See also the maps of $\langle b_j \rangle$ in Appendix B.

As a final remark for the transient, we see that $\tilde t \approx { O}(1)$ corresponds also to the typical time scales of the fluctuations of $N_0$ and $C$. It is actually quite natural that the finite-size inhomogeneous dynamical system defined by Eq.~(\ref{eq:tdG}) shows such fluctuations.  We show the effects of changing the system size in the Appendix A.

\begin{figure}
 \centering
 \includegraphics[width=0.48\textwidth,clip=true]{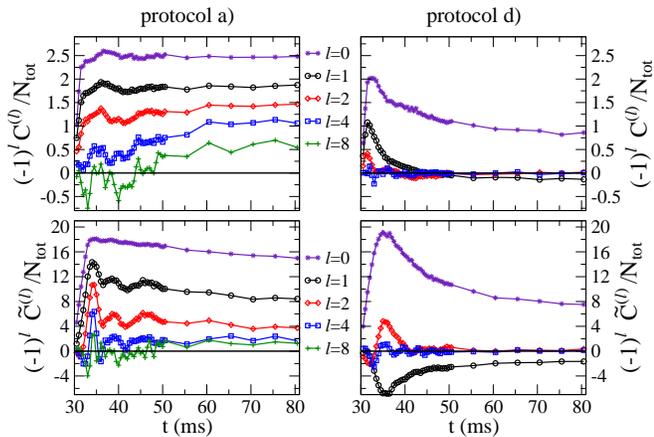}
 \caption{(color online) \label{fig:CCtilde} Coherence functions $(-1)^l C^{(l)}$ (top) and $(-1)^l \tilde C^{(l)}$ (bottom) for the protocols a) (left) and d) (right) as a function of time $t$ for different spacings $l=0,1,2,4,8$. }
\end{figure}

Now we focus on the coherence of the superfluid order parameter in the final optical lattice. We show examples of snapshots of the field $\langle b_j \rangle$ in Appendix B. To analyze the spatial distributions, we define (equal-time) cloud averages of coherences at longer range,
\begin{equation}
 C^{(l)}(t) := \sum_j \sum_{ \delta} \langle b_j^\dagger(t) \rangle \langle b^{\phantom{\dagger}}_{j+ l \delta}(t) \rangle, \label{eq:def:Cn}
\end{equation}
where $\delta$ runs over the four nearest-neighbor directions. Therefore, $C^{(0)} = 4 N_0$ and $C^{(1)} = C$. In addition, we also define 
\begin{equation}
 \tilde C^{(l)}(t) := \sum_j \sum_{\delta} \frac{\langle  b_j^\dagger(t) \rangle}{\sqrt{n_j(t)}} \frac{\langle  b^{\phantom{\dagger}}_{j+ l \delta}(t) \rangle}{\sqrt{n_{j+l\delta}(t) } } \label{eq:def:Cntilde}
\end{equation}
using the density-normalized order parameter, which is more sensitive to the edge of the cloud. Note that $| \langle b_j(t) \rangle|/\sqrt{n_j(t)} \to 0$ for $|\mathbf{r}_j|\to \infty$; therefore, the sum converges.
We emphasize that these quantities are defined through \emph{averages over the cloud}, as quantum-mechanical expectation values enter the definitions explicitly. 
The coherences are shown in Fig.~\ref{fig:CCtilde} as a function of time. Since $C$ is negative ($C^{(l=1)} < 0$) we display $(-1)^l C^{(l)}$ in Fig.~\ref{fig:CCtilde}. 

First we focus on protocol a). The spatial decay of the coherences can be used to define a characteristic length.
First, we average the data points for each $l$ for $t\geq 60$ ms, yielding a mean value $\bar C^{(l)}$ and a variance $\sigma^2(C^{(l)})$. Then we perform a least-squares fit with an exponential function $A \exp(-l/l_c)$ to the data $\lbrace l,(-1)^l \bar C^{(l)} \rbrace_{l=0,1,2,4,8}$ weighted with $\sigma^{-2}(C^{(l)}) $. This fit yields a characteristic length $l_c = 3.90 \pm 1.18 $, which is accidentally consistent with the estimate of the coherence length in Ref.~\cite{exp-negT}.
The main qualitative difference between $C$ and $\tilde C$ is that the density-normalized coherences do not seem to be stationary. We have to keep in mind that $\tilde C$ is more sensitive to the dilute edges of the cloud, where approaching stationary values is much slower. This is actually similar to the fermionic case ( cf. Fig. 3 of Ref.~\cite{negT}): the edges still show significant deviations even when the local inverse temperatures are more or less homogeneous in the middle of the cloud. 

In the case of protocol d), longer-range coherences decay rapidly as a function of time following a transient behavior. In stark contrast to protocol a), where both $C^{l=1} < 0$ and $\tilde C^{l=1} < 0$, for the trapping case d) the density-normalized coherences are actually \emph{positive}, $\tilde C^{l=1} > 0$. 
Microscopically, this can be explained assuming that the core and edge are subject to competing processes during the transient, but eventually, the dynamics that governs the tails of the cloud wins. The process in the middle is related to the dominance of the attractive interaction over the harmonic potential, and therefore anticoherence tries to develop. On the other hand, in the edge of the cloud, the harmonic potential is stronger and the interaction is effectively weaker due to the low densities. Thus, in the edge the corresponding dynamics prefers positive coherences dictated by the trapping potential.
After the cloud expands from its initial compressed state (see Fig.~\ref{fig:cd}), even the coherence becomes positive, $C>0$.

While the finite but long observed coherence lifetimes~\cite{exp-negT} for antitrapping potentials and attractive interactions in the experiments are most likely due to loss processes~\cite{losses} which are not taken into account by Eqs.~(\ref{eq:BHM}) and (\ref{eq:tdG}), the experimental fact that coherence is lost much faster for protocols similar to c), d), or e) is captured by the present approach. Unfortunately, based on the current simulations we cannot extract a ``mean-field'' coherence lifetime $\tau_{\rm MF}$, which should depend on the system size $N_{\rm tot}$ as well as on the microscopic parameters, most importantly on $V_0$. If the experimental losses are characterized by lifetimes $\tau_{\rm 1B}$ and $\tau_{\rm MB}$ referring to one-body and many-body loss processes, respectively, the coherence lifetime is expected to be given by $\bar \tau^{-1} \approx \tau_{\rm MF}^{-1} + \tau_{\rm 1B}^{-1} + \tau_{\rm MB}^{-1}  $, as the different processes are independent  ``channels'' for decoherence. 
Note that the shortest lifetime dominates; therefore, the experimental lifetime can be qualitatively explained provided that  $\tau_{\rm MF}( U<0, V_0<0 ) \gg  \tau_{\rm 1B}, \tau_{\rm MB} $ (compatible with the macroscopic stability) while $\tau_{\rm MF}( U<0, V_0 \to + \infty ) \to 0$.

\section{Conclusions}

We used time-dependent Gutzwiller mean-field theory to describe the dynamics of bosons which are subject to various protocols with the aim of achieving $T<0$ with ultracold atoms in optical lattices. We found that these numerical results qualitatively agree with the experimental findings. Most importantly, after reversing the signs of both the interactions between the atoms and the external harmonic potential, the system relaxes to a macroscopically stationary state where the BEC shows up at finite momenta. This novel BEC pattern can be directly seen in the TOF images. The simple time-dependent mean-field theory also seems to capture the experimental observation that the combination of an attractive final interaction and a trapping final potential leads to a rapid decay in TOF images. 

Furthermore, the mean-field picture obtained by the time-dependent Gutzwiller approach provides some insight about the microscopic dynamics. Whether this picture is correct can only be answered by further studies. Some of the quantities calculated from the GA could be accessed experimentally by taking \textit{in situ} images of the atomic density in real space, which could be used to measure the time evolution of the cloud radius $R$ or possibly also the Mott radius $R_{\rm Mott}$. Other quantities, especially the density-normalized coherence $\tilde C^{(l=1)}$, can at the moment only be compared to other numerical simulations.

Despite that we found qualitative similarities for some quantities between the numerics and the experiments, one has to be careful about
quantitative comparisons. The experiment operates on a three-dimensional cloud, which is too large for the current numerics. The evaluation of microscopic parameters can also be improved by using the integrals of the numerically exact Wannier functions. Additionally, thermodynamic quantities, like entropy or temperature, cannot be introduced in this GA in a natural way. Addressing these and further restrictions are beyond the scope of the present work and left for future studies.

\textbf{Acknowledgments:} I thank Luis Santos, Mattia Jona Lasinio, Hendrik Weimer, Achim Rosch, and especially Ulrich Schneider for discussions. This work has been supported financially by the DFG.

\appendix
\section{Finite size scaling in the fluctuations}

\begin{figure}[h]
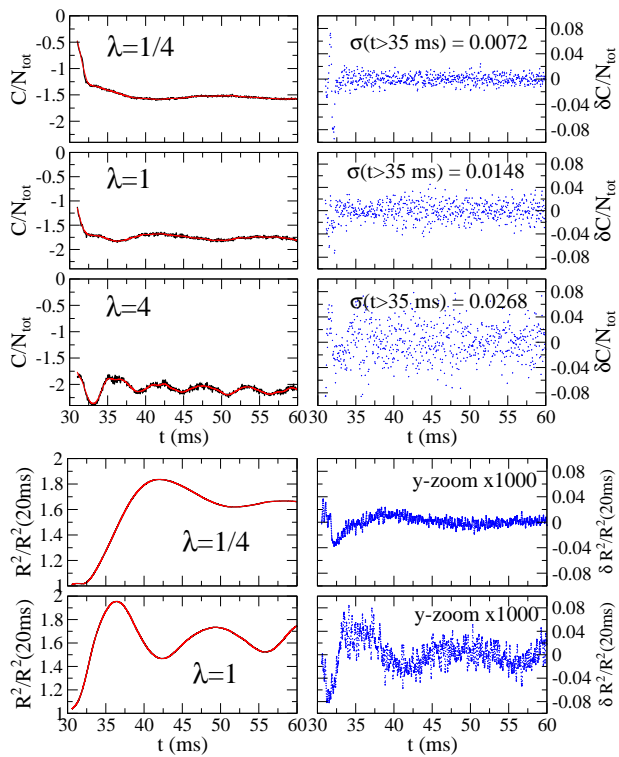

 \centering
 \includegraphics[width=0.45\textwidth,clip=true]{finitesize.eps}
 
 \includegraphics[width=0.45\textwidth,clip=true]{finitesizeR.eps}
 \caption{(color online) \label{fig:finitesize} Finite-size effects in different quantities as function of time. The left figures show the ``raw'' data (thick black line) and a smoothed curve (thin red line). The right figures show the difference. For the coherences $C$, averages of 1000 raw data points were used for smoothing. Due to the larger oscillation amplitudes in $R^2$, only 100 raw data points were taken. }
\end{figure}

To investigate the effects of the system size on the noise of the data, we performed simulations with different atom numbers. For two-dimensional systems in the presence of a harmonic potential, the thermodynamic limit is the limits $N_{\rm tot} \to \infty$ and $|V_0| \to 0$ taken simultaneously while the ``compression,'' $V_0 N_{\rm tot}$, is kept fixed. For the tightly compressed initial system this latter condition is achieved by keeping the central chemical potential $\mu_0$ fixed. Therefore, we introduce a scaling variable $\lambda$ and compare systems with different potential strengths
\begin{equation}
 V_0 \to V_0' = \lambda V_0 \;,
\end{equation}
which implies atom numbers $N_{\rm tot}' \approx \lambda^{-1} N_{\rm tot}$. Note that it is expected that the frequency of oscillations in the harmonic trap also scales as $\sim \sqrt{|V_0'|} \sim \sqrt{\lambda}$. We used the actual values $\lambda=1/4,1,$ and $4$, where $\lambda =1$ corresponds to the system used in the main text, with $N_{\rm tot} \approx 1920$. In Fig.~\ref{fig:finitesize} we show the coherences $C$ and the system radius squared, $R^2$, for different values of $\lambda$.

To extract the noise, we apply a simple data-smoothing procedure based on averaging a certain amount of data points. For the case of the coherences, we used averages of 1000 ``raw'' data points. The difference $\delta C$ between the raw data and the smooth curve is also shown in Fig.~\ref{fig:finitesize}. The standard deviation $\sigma$ of $\delta C$ for $t>35$ ms seems to scale approximately with $\sim \sqrt{\lambda} \sim 1/\sqrt{N_{\rm tot}'}$. 
In addition, we also show the fluctuations of the radius square, $\delta(R^2)$, which are orders of magnitude weaker than $\delta C$.
Since the oscillation amplitude is much larger, we had to use fewer  raw data points, 100, for the average; however, it should only account for an approximately $\sqrt{10}$-fold decrease in the relative fluctuation strength.
Note that the noise of the radius, $\delta R$, is even more suppressed since $\delta(R^2) \approx 2 R \, \delta R $.

To understand why only $N_0$ and $C$ show visible fluctuations, we have to keep in mind that $N_{\rm tot}$ is conserved and that $R$ (or $R^2$) depends only on the absolute values $|f_m(j,t)|$, in contrast to $N_0$ and $C$ which also depend explicitly on the phases, $\arg( f_m(j,t) )$. 
Note that, due to the normalization constraints, $|f_m(j,t)|$ is confined to a smaller interval 
than the phases which can take values up to $2\pi$.

\section{Order parameter maps}

\begin{figure*}
 \centering
 \includegraphics[width=0.95\textwidth,clip=true]{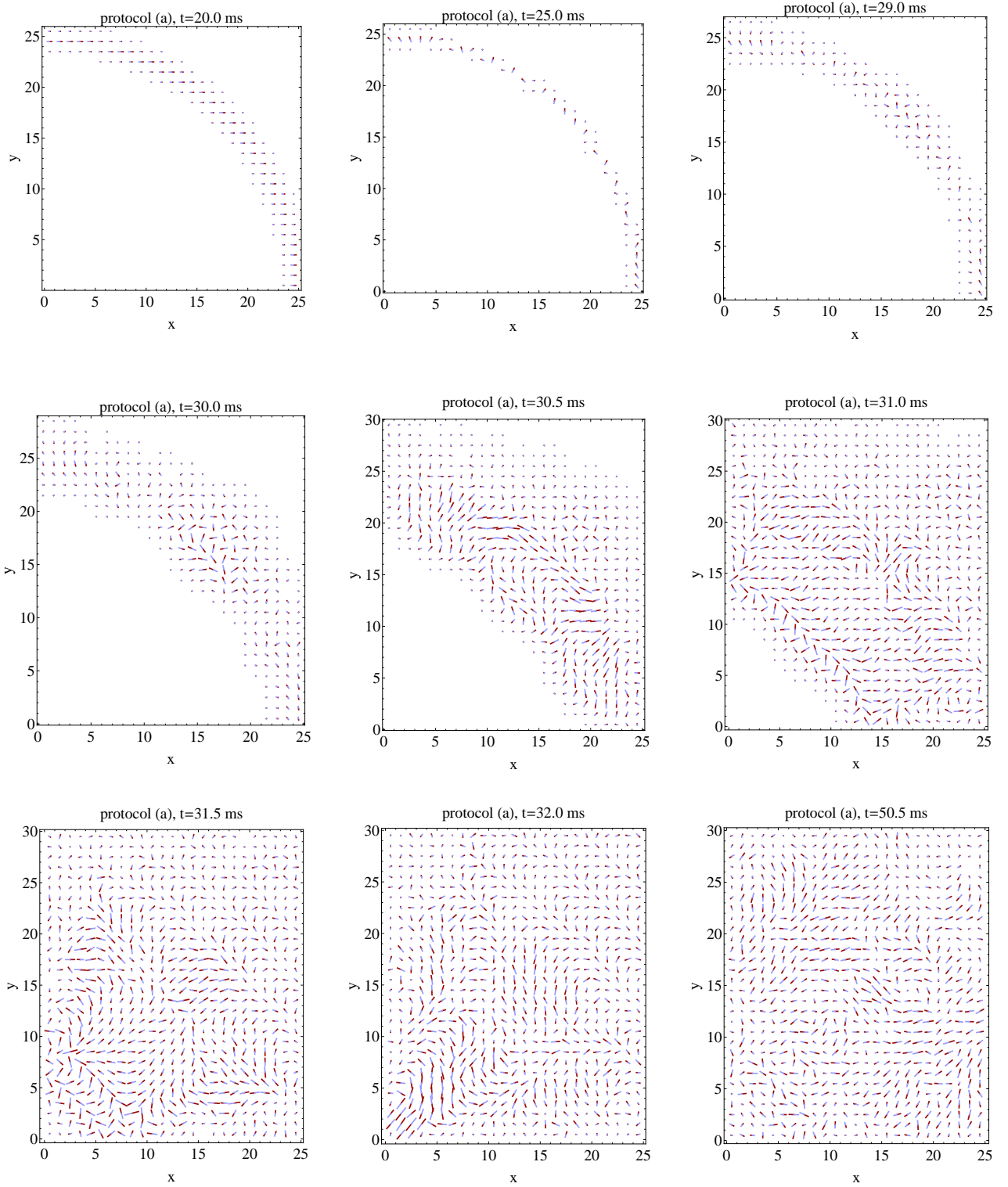}
 \caption{(color online) \label{fig:maps} Snapshots of the field $\langle  b_j(t) \rangle$. At each site $\mathbf{r}_j=(x_j,y_j)$ the order parameter is represented by a compass. The length of the compass equals $|\langle  b_j(t) \rangle|$, while the orientation (light blue to dark red) corresponds to ${\rm arg}[ \langle  b_j(t) \rangle ]$. Note how the Mott-insulating core melts during the transient. In the stationary regime, for the protocol z) (cf. Fig.~\ref{fig:maps0}), regions become mostly aligned, while for the protocol a) alignments mostly become antiparallel. }
\end{figure*}

\begin{figure*}
 \centering
 \includegraphics[width=0.95\textwidth,clip=true]{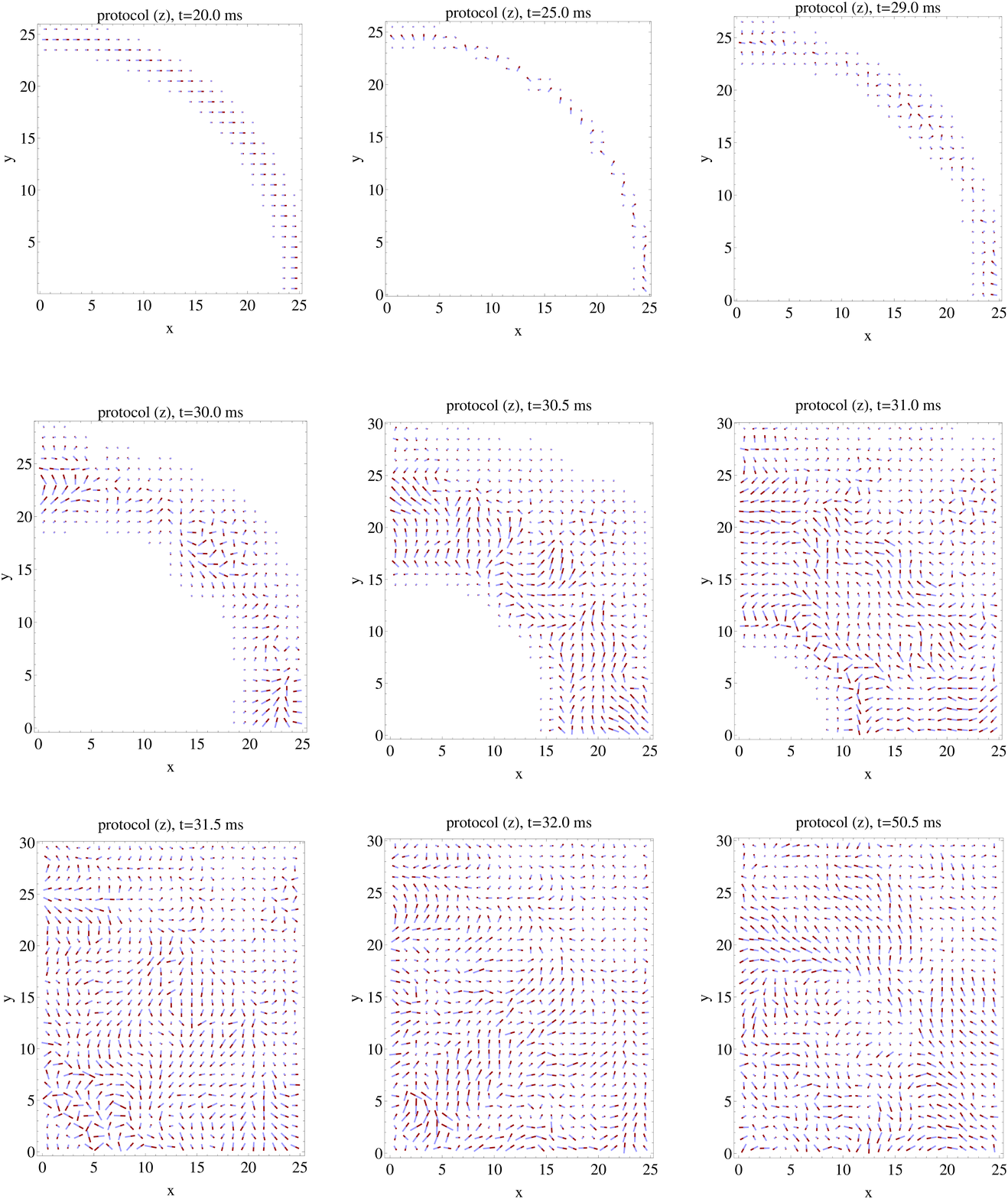}
 \caption{(color online) \label{fig:maps0} Snapshots of the field $\langle  b_j(t) \rangle$. At each site $\mathbf{r}_j=(x_j,y_j)$ the order parameter is represented by a compass. The length of the compass equals $|\langle  b_j(t) \rangle|$, while the orientation (light blue to dark red) corresponds to ${\rm arg}[ \langle  b_j(t) \rangle ]$. Note how the Mott-insulating core melts during the transient. In the stationary regime, for the protocol z), regions become mostly aligned, while for the protocol a) (cf. Fig.~\ref{fig:maps}) alignments mostly become antiparallel. }
\end{figure*}

We show order parameter snapshots of $\langle b_j(t) \rangle$ in Figs.~\ref{fig:maps} and \ref{fig:maps0}. This helps to visualize how the Mott-insulating core melts and also to display the correlated regions for the stationary regime. For any site $j$ the complex number corresponding to $\langle b_j(t) \rangle$ can be simply represented by a two-dimensional vector. Initially, there is a completely coherent, but weak, condensate
around a Mott region, at a radius $R_{\rm Mott} \approx \sqrt{2} R(20\; {\rm ms})$, which follows from the tight confinement. As the lattice depth is decreased after $t=28$ ms, the Mott region melts away gradually. It seems that the local condensate fraction $|\langle b_j(t) \rangle|^2$ increases first and only afterwards do the phases start to align.

\begin{thebibliography}{99}
\bibitem{losses} The long experimental coherence lifetimes, exceeding 600 ms, imply
		 that such loss processes only weakly affect the cloud dynamics 
		 for the current simulation times.
\end{thebibliography}
\end{document}